\documentstyle[12pt,epsf]{article}

\def\beq{\begin{equation}}
\def\eeq{\end{equation}}
\def\barr{\begin{eqnarray}}
\def\earr{\end{eqnarray}}
\def\dag{\dagger}
\def\b{\bigskip}
\def\tr{{\rm tr}}
\def\l{\left(}
\def\r{\right)}

\begin{document}

\input FEYNMAN
\baselineskip=0.35in
\title{Renormalization In Coupled-Abelian Self-Dual Chern-Simons Models}

\author{\normalsize{Gerald Dunne and Theodore Hall} \\
\normalsize{Department of Physics}\\
\normalsize{University of Connecticut}\\
\normalsize{Storrs, CT 06269 USA}\\
   \\
\normalsize{dunne@hep.phys.uconn.edu}\\
\normalsize{tedhall@main.phys.uconn.edu} \\}

\date{}

\maketitle

\begin{abstract}
An algebraic restriction of the nonabelian self-dual Chern-Simons-Higgs systems
leads to coupled-abelian self-dual models with intricate mass spectra. The
vacua are characterized by embeddings of $SU(2)$ into the gauge algebra; and in
the broken phases, the gauge and real scalar masses are related to the
exponents of the gauge algebra. In this paper we compute the gauge-gauge-Higgs
couplings in the broken phases and use this to compute the finite
renormalizations of the Chern-Simons coefficient in the various vacua.
\end{abstract}

Chern-Simons theories are well-known to exhibit interesting topological and
renormalization properties. A Dirac-style quantum consistency condition leads
to an integer quantization condition ($\kappa=\frac{1}{4\pi}({\rm integer})$)
on the coefficient $\kappa$ of the Chern-Simons term in the Lagrangian
\cite{deser}. Such a consistency condition is remarkably robust under
renormalization; for example, in an $SU(N)$ Chern-Simons-Yang-Mills theory, the
bare Chern-Simons coefficient $\kappa$ receives a (finite) additive
renormalization shift which is an {\bf integer} multiple of $\frac{1}{4\pi}$:
$\kappa\to\kappa+\frac{N}{4\pi}$ \cite{pisarski}. When charged matter fields
are coupled to the Chern-Simons gauge fields, this issue becomes more involved
due to spontaneous symmetry breaking effects. Direct diagrammatic analyses have
led to the conclusion that when the symmetry is broken completely the
diagrammatic shift in $\kappa$ is some complicated noninteger (dimensionless)
combination of the Chern-Simons mass scale and the Higgs mass scale
\cite{khlebnikov}; but when the symmetry is only partially broken, leaving a
residual nonabelian symmetry in the broken phase, this dimensionless shift for
the broken generators is once again an integer multiple of $\frac{1}{4\pi}$
(with the integer being the Coxeter number of the residual nonabelian symmetry
group) \cite{lusheng}. A deeper way to understand these results is through a
generalization of the Coleman-Hill theorem \cite{coleman} to incorporate
symmetry breaking systems \cite{khare}. It is believed that the complicated
shifts of $\kappa$ derived in the diagrammatic approach actually reflect the
appearance of gauge invariant terms in the effective action which mimic a
Chern-Simons term at large distances as the Higgs field tends to its nonzero
vacuum expectation value \cite{khlebnikov,khare}. This is an appealing and
consistent picture, but it has only been explicitly demonstrated in abelian
theories \cite{khare}, not in nonabelian theories.

Another important feature of Chern-Simons theories is that they admit a
self-dual form of matter-gauge theory. Such systems have a Bogomol'nyi lower
bound for the energy, which is saturated by solutions to a set of first-order
self-duality equations \cite{hong}. These systems have a special sixth-order
Higgs potential (renormalizable in $2+1$ dimensions), and a consequence of the
particular form of this potential is that the Chern-Simons mass scale is a
multiple of the Higgs mass scale. In the abelian self-dual Chern-Simons system
in the broken phase, the Higgs and gauge masses are in fact equal and a
remarkable consequence of this is that the one-loop renormalization shift of
the Chern-Simons parameter is just $\frac{1}{4\pi}$ \cite{kao2}. In this case
the source of the integer-quantized shift is the self-duality of the model,
rather than the existence of a nonabelian symmetry in the broken phase. The
purpose of this paper is to investigate whether or not such a quantized shift
occurs in the more intricate coupled-abelian models studied in \cite{dunne2},
for which there are many inequivalent broken vacua, and for which the gauge and
Higgs mass spectra reveal surprising degeneracy patterns. These coupled-abelian
theories are based on a nonabelian gauge algebra ${\cal G}$, but are
algebraically restricted (in a manner analogous to Toda theories) so that the
coupling between the fields involves the Cartan matrix of ${\cal G}$. For
simplicity we restrict our attention to simply-laced algebras, and we will
often concentrate more specifically on $SU(N)$.

The nonabelian self-dual Chern-Simons-Higgs theory \cite{klee,kao1,dunne1} is
described by the following Lagrange density (in $2+1$ dimensional spacetime)
\beq
{\cal L}=-\tr\l\l D_\mu\phi\r^\dag D^\mu\phi \r -\kappa
\epsilon^{\mu\nu\rho} \tr\l \partial_\mu A_\nu A_\rho + {2 \over 3}
A_\mu A_\nu A_\rho \r - V\l\phi,\phi^\dag\r
\label{lag}
\eeq
where $V(\phi,\phi^\dag)$ is the sixth-order self-dual potential

\beq
V(\phi,\phi^\dag)={1\over 4\kappa^2}\tr\left\{
\l[\;[\;\phi,\phi^\dag\;],\phi\;]-
v^2 \phi \r^\dag \l [\:[\; \phi,\phi^\dag\;],\phi\;]-v^2 \phi \r\right\}.
\label{pot}
\eeq
The space-time metric is taken to be $g_{\mu\nu}=diag(-1,1,1)$, and we work
with adjoint coupling so that the covariant derivative takes the form
$D_\mu=\partial_\mu+[A_\mu,\ ]$. The trace runs over the

finite-dimensional representation of the compact simple Lie algebra $\cal G$
to which the gauge field $A_\mu$ and the charged matter fields $\phi$
and $\phi^\dag$ belong.

With the self-dual potential (\ref{pot}) there exists a Bogomol'nyi lower bound
on the energy density which is saturated by solutions to the (relativistic)
self-dual Chern-Simons equations
\barr
D_-\phi&=&0\nonumber \\
F_{+-}&=&{1\over \kappa^2}[\;(v^2\phi-[\;[\;\phi,\phi^\dag\;],
\phi\;]\;),\phi^\dagger\;]
\label{sd}
\earr
where $D_-\equiv D_1-iD_2$. Solutions to these self-duality equations fall into
classes characterized by the asymptotic values of the scalar fields,
corresponding to the gauge inequivalent vacua of the self-dual potential
(\ref{pot}). These (degenerate) vacua are determined by the algebraic condition
\beq
[\;[\;\phi,\phi^\dag\;],\phi\;]=v^2\phi
\label{embedding}
\eeq
Identifying $\frac{1}{|v|}\phi$ with $J_+$ (and $\frac{1}{|v|}\phi^\dagger$
with $J_-$), this vacuum condition is equivalent to the $SU(2)$ commutation
relations. Thus, the vacua are classified by the
inequivalent embeddings of $SU(2)$ into the gauge algebra ${\cal G}$; such
embeddings are important in Lie algebra theory \cite{dynkin}, and it is
interesting to note that they have also featured prominently in other
well-known self-dual gauge theories \cite{sorba}. For $SU(N)$, the number of
inequivalent vacua (including the trivial $\phi=0$ one) is equal to the number,
$p(N)$, of partitions of $N$.

The coupled-abelian models to be considered in this paper are obtained from
(\ref{lag},\ref{pot},\ref{sd}) by restricting the fields according to the
algebraic ansatz
\barr
\phi  = \sum_{a=1}^r \phi^a E_a  \hskip 2cm
A_{\mu}= i \sum_{a=1}^r A_{\mu}^a H_a
\label{ansatz}
\earr
where r is the rank of ${\cal G}$. We work in a Chevalley basis for the Lie
algebra, with $H_a$ and $E_a$ being the Cartan subalgebra and simple root step
operators respectively. These generators satisfy the commutation and
normalization conditions

\beq\begin{array}{llll}
[H_a,H_b]&=0 \hskip .5in & \tr (H_aH_b)&= K_{ab} \cr
[E_a,E_{-b}]&= \delta_{ab} H_b \hskip .5in &
\tr(H_a E_b)&=0 \cr
[H_a,E_{\pm b}]&= \pm K_{ab} E_{\pm b} \hskip .5in &
\tr (E_a E_{-b})&= \delta_{ab}
\end{array}
\label{chevalley}
\eeq
where $E_{-a}=E_a^\dagger$, and the Cartan matrix $K$ expresses the inner
products of the simple roots $\vec{\alpha}^{(a)}$ (normalized to have length
$\sqrt{2}$)

\beq
K_{ab}=  \vec{\alpha}^{(a)} \cdot \vec{\alpha}^{(b)}\qquad (a,b=1\dots r)
\label{cartan}
\eeq
As shown in \cite{dunne2}, the algebraic restriction (\ref{ansatz}) is
self-consistent. The self-dual potential (\ref{pot}) is now expressed in terms
of $r$ complex scalar fields $\phi^a$
\barr
V={v^4\over 4\kappa^2} \sum_{a=1}^r | \phi^a |^2 -{v^2 \over
2 \kappa^2}\sum_{a=1}^r\sum_{b=1}^r|\phi^a|^2 K_{ab} |\phi^b|^2 +{1\over
4\kappa^2}\sum_{a=1}^r\sum_{b=1}^r\sum_{c=1}^r |\phi^a|^2 K_{ab} |\phi^b|^2
K_{bc}|\phi^c|^2
\label{r-pot}
\earr

\begin{figure}[h]
\epsffile{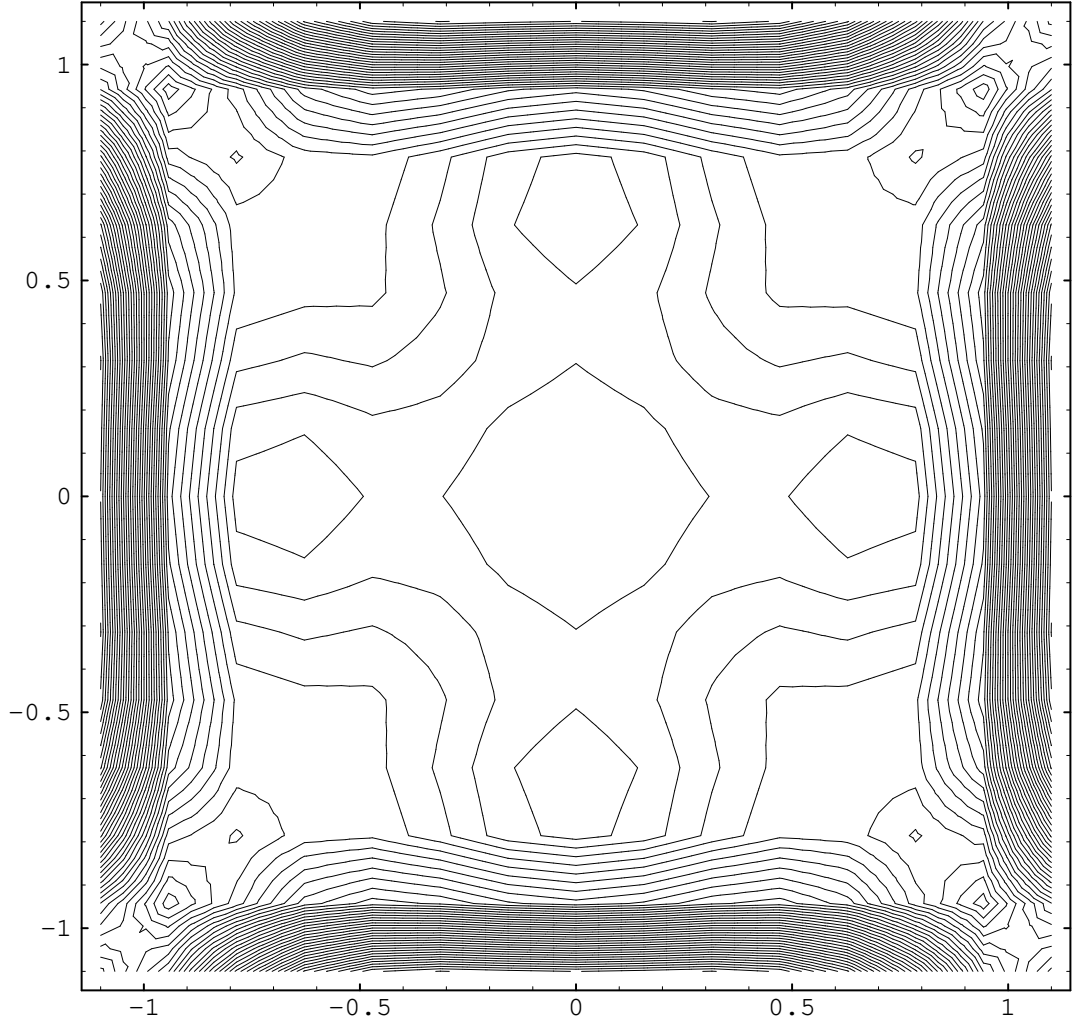}
\caption{Contour plot of the self-dual potential $V(\phi^1,\phi^2)$ for the
restricted $SU(3)$ model. Note the three classes of inequivalent vacua: (i) at
$(\phi^1,\phi^2)=(0,0)$; (ii) at $(\phi^1,\phi^2)=(\pm 1/\protect{\sqrt{2}},0)$
and $(0,\pm 1/\protect{\sqrt{2}})$; and (iii) at $(\phi^1,\phi^2)=(\pm 1,\pm
1)$ and $(\pm 1,\mp 1)$.}
\label{potential}
\end{figure}

For $SU(2)$ this reduces to $V=\frac{1}{4\kappa^2}|\phi|^2(2|\phi|^2-v^2)^2$,
which is just the self-dual abelian potential studied in \cite{hong}, and which
has two inequivalent vacua. For $SU(3)$,
$K_{ab}=\left(\matrix{2&-1\cr-1&2}\right)$, and the self-dual potential is
$V=\frac{1}{4\kappa^2}(4(|\phi^1|^6+|\phi^2|^6)-3(|\phi^1|^4|\phi^2|^2
+|\phi^1|^2|\phi^2|^4) -4v^2(|\phi^1|^4-|\phi^1|^2|\phi^2|^2 +|\phi^2|^4)
+v^4(|\phi^1|^2 +|\phi^2|^2))$. A contour plot of this function
$V(\phi^1,\phi^2)$ is shown in Figure \ref{potential}. Note the presence of the
three inequivalent vacua: $(|\phi^1|,|\phi^2|)=(0,0)$,
$(|\phi^1|,|\phi^2|)=\frac{|v|}{\sqrt{2}}(1,0)$, and
$(|\phi^1|,|\phi^2|)=|v|(1,1)$. [The minimum at
$(|\phi^1|,|\phi^2|)=\frac{|v|}{\sqrt{2}}(0,1)$ is equivalent to the one at
$(|\phi^1|,|\phi^2|)=\frac{|v|}{\sqrt{2}}(1,0)$.] We note that the algebraic
restriction (\ref{ansatz}) is consistent with the vacuum condition
(\ref{embedding}), since a representative $\phi_{(0)}$ from each equivalence
class of minima may be expanded as
\beq
\phi_{(0)}=|v| \sum_{a=1}^r \phi_{(0)}^a E_a
\label{min}
\eeq
where the $\phi_{(0)}^a$ are some numerical coefficients. For example, the
``maximal embedding vacuum'' is defined by taking $(\phi_{(0)}^a)^2$ to be the
coefficients of the decomposition of (half) the sum of positive roots, in terms
of the simple roots:
$\vec{\rho}\equiv\frac{1}{2}\sum_{\alpha>0}\vec{\alpha}=\sum_{a=1}^r
(\phi_{(0)}^ a)^2\vec{\alpha}^{(a)}$. For $SU(N)$ this means
$\phi_{(0)}^a=\sqrt{a(N-a)/2}$,
$a=1\dots N-1$. On the other hand, the simplest nontrivial vacuum solution to
(\ref{embedding}) is obtained by taking $\phi_{(0)}$ to involve just one step
operator, which (up to equivalence) we can choose to be $E_1$:
\beq
\phi_{(0)}=\frac{|v|}{\sqrt{2}}E_1
\label{simp}
\eeq
All other vacua are also specified by a decomposition of the form in
(\ref{min}), and may be classified in terms of the Dynkin diagrams of the
various subalgebras of ${\cal G}$ \cite{dunne1}. Since these vacua are, in
turn, characterized by the maximal embedding vacua for the various subalgebras
so obtained, we can limit our attention to the maximal embedding vacuum. It is
a straightforward matter to pass to the other vacua.

Given $\phi_{(0)}$, the gauge and scalar mass spectra in that vacuum are
determined by shifting $\phi\rightarrow \phi+\phi_{(0)}$, and keeping terms in
the Lagrangian that are quadratic in $A_\mu$ and $\phi$. The quadratic gauge
part of the Lagrangian is
\barr
{\cal L}_{quad}^{gauge}= -{\kappa} \epsilon^{\mu\nu\rho} \:\tr ( \partial_\mu
A_\nu A_\rho ) - \:\tr \l [\phi_{(0)},A_\mu]^\dag [\phi_{(0)},A^\mu ] \r
\label{gauge-quad}
\earr
Thus the gauge mass matrix is
\barr
M_{gauge}=\frac{1}{2\kappa}\left(ad(\phi_{(0)}^\dagger) ad(\phi_{(0)})+
ad(\phi_{(0)})ad(\phi_{(0)}^\dagger)\right)
=\frac{v^2}{2\kappa}\l J_+J_-+J_-J_+\r
=\frac{v^2}{2\kappa}{\cal C}
\label{gauge-mass}
\earr
where ${\cal C}$ is the $SU(2)$ quadratic Casimir for the particular embedding
(we have also used the fact that $J_3 A\equiv 0$, which follows from the
algebraic restriction of the gauge field to the Cartan subalgebra in
(\ref{ansatz})). For the maximal embedding case, the adjoint action of $SU(2)$
decomposes ${\cal G}$ into $r$ subblocks, of spin $s_a$, where the integers
$s_a$ are called the {\it exponents} of ${\cal G}$ \cite{dynkin}. Thus, in the
maximal embedding vacuum all $r$ gauge fields acquire nonzero mass, with masses
\beq
m_a=m\, s_a(s_a+1)\hskip 2cm a=1\dots r,
\label{exp}
\eeq
where $m=\frac{v^2}{2\kappa}$ is a common mass scale, which is equal to the
scalar mass in the unbroken phase. For $SU(N)$ the exponents are the integers:
$1,2,\dots , N-1$.

The scalar masses are obtained from the quadratic part of $V(\phi+\phi_{(0)})$.
In the maximal embedding vacuum this leads to a mass (squared) matrix
\barr
M_{scalar}^2= \frac{1}{\kappa^2} \left(ad(\phi_{(0)}^\dagger)
ad(\phi_{(0)})\right)^2=m^2 {\cal C}^2
\label{scalar-mass}
\earr
where we have used the fact that $J_3\phi=\phi$, which follows from the
algebraic decompostion of the $\phi$ field in (\ref{ansatz}). Hence, in the
maximal embedding vacuum, there are $r$ nonzero scalar masses, and these are
{\it equal} to the gauge masses in (\ref{exp}).

These mass spectra refer to {\it classical} properties of the self-dual
Chern-Simons system (\ref{lag},\ref{ansatz}), and the degeneracy of the gauge
and scalar masses is a reflection of the underlying $N=2$ supersymmetry of
these self-dual Chern-Simons models \cite{susy}. However, the masses are also
important for quantum effects. In particular, consider the one-loop
renormalization of the Chern-Simons coupling parameter $\kappa$. It is by now a
standard diagrammatic computation to show that $\kappa$ receives a finite
additive shift (in Landau gauge) obtained from the zero (external) momentum
limit of the parity-odd part of the gauge self-energy diagram shown in Figure
2.

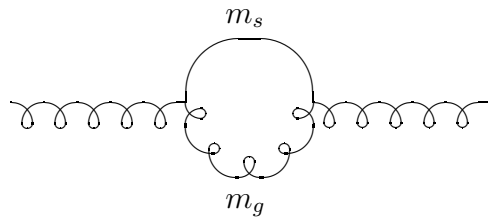
\begin{figure}[h]
\begin{center}
\begin{picture}(6000,5000)
\frontstemmed\drawline\gluon[\W\FLIPPEDFLAT](10000,0)[5]
\drawloop\gluon[\S 5](\gluonbackx,\gluonbacky)
\put(1000,0){\oval(4800,4800)[t]}
\frontstemmed\drawline\gluon[\W\FLIPPEDFLAT](\gluonbackx,\gluonbacky)[5]
\startphantom
\frontstemmed\drawline\gluon[\E\FLIPPEDFLAT](\gluonbackx,\gluonbacky)[5]
\stopphantom
\put(100,3000){$m_s$}
\put(100,-4000){$m_g$}
\end{picture}
\end{center}
\vskip .3in
\caption{Gauge self-energy arising from gauge-scalar interaction.}
\end{figure}

The shift $\Delta\kappa$ is given by

\barr
\Delta\kappa&=& {i \over2} \lim_{p^2\to 0}{p_\gamma \over p^2}
\epsilon^{\gamma\mu\nu} \Pi_{\mu\nu} (p) \cr
&=& -{i \over 2 \pi^3} {\Gamma^2\over \kappa}\lim_{p^2 \to 0} \int d^3q \;
{p \cdot q \over p^2((p-q)^2 -m_s^2 ) (q^2 - m_g^2 ) }\cr
&=&{1 \over 4\pi} {4 \over 3}  {\Gamma^2 \over \kappa}
{m_s + 2m_g \over (m_s+m_g)^2}
\label{shift}
\earr
where $\Gamma$ is the vertex factor giving the weight of the coupling between
the gauge and scalar fields of mass $m_g$ and $m_s$ respectively. In the broken
phase of the abelian system, the gauge and scalar masses are degenerate, and
the gauge-gauge-scalar vertex factor is $\Gamma^2=|\kappa | m_s=|\kappa | m_g$,
so that $\Delta\kappa=sign(\kappa)\frac{1}{4\pi}$ \cite{kao2}. To determine the
analogous renormalization shifts for the coupled-abelian system with self-dual
potential (\ref{r-pot}) one needs, in addition to the masses (\ref{exp}) of the
gauge and scalar fields, the three-point gauge-gauge-scalar coupling factors,
which are determined by the cubic term
\beq
\tr\l[\phi^\dagger,[A^\mu ,\phi_{(0)}]]A_\mu^\dag\r+ \tr\l[\phi_{(0)}^\dagger,
[A^\mu ,\phi]]A_\mu^\dag\r
\label{cubic}
\eeq
However, the fields $A$ and $\phi$ must first be decomposed in terms of the
basis of {\it physical} fields which diagonalize the respective mass matrices.
This requires knowledge of not just the {\it eigenvalues} of the mass matrices,
but also of the corresponding {\it eigenvectors}. This may be achieved by the
following construction. Define an $r\times r$ diagonal matrix
$D=diag(\phi_{(0)}^1, \phi_{(0)}^2, \dots, \phi_{(0)}^r)$, where the diagonal
entries $\phi_{(0)}^a$ are the coefficients of the decomposition (\ref{min}) of
the vacuum solution $\phi_{(0)}$. (Note that we are considering the maximal
embedding vacuum, so that all the $\phi_{(0)}^a$ are nonzero). It is then
straightforward to check that the gauge and scalar mass matrices are given by
\barr
M_{gauge}=2m \,D^2\,K\hskip 2cm
M^2_{scalar}=4m^2 \,\left(D\, K\, D\right)^2
\label{masses}
\earr
where $K$ is the Cartan matrix (\ref{cartan}). Now define the eigenvectors
$\vec{\lambda}^{(a)}$ and $\vec{\mu}^{(a)}$ of the matrices $D^2 K$ and $DKD$
respectively:
\barr
(D^2K) \vec{\lambda}^{(a)}=\frac{1}{2}s_a(s_a+1)\vec{\lambda}^{(a)}\hskip 2cm
(DKD)\vec{\mu}^{(a)}=\frac{1}{2}s_a(s_a+1)\vec{\mu}^{(a)}
\label{eigenvectors}
\earr
These eigenvectors can be normalized as
\beq
\vec{\lambda}^{(a)T}K\vec{\lambda}^{(b)}=\delta^{ab}\hskip 2cm
\vec{\mu}^{(a)T}\vec{\mu}^{(b)}=\delta^{ab}
\label{norm}
\eeq
Given these eigenvectors, we now decompose the gauge and scalar fields $A$ and
$\phi$ as (we suppress the spacetime index on the gauge field as it is not
important here)
\barr
A&=&{i\over \sqrt{2}} \sum_{a=1}^r
A^a_{phys}\left(\vec{\lambda}^{(a)}\cdot\vec{H}\right)\cr
\phi&= &\frac{1} {\sqrt{2}} \sum_{a=1}^r\phi_{phys}^a
\left(\vec{\mu}^{(a)}\cdot\vec {E}\right)
\label{physical}
\earr
Then the fields $A_{phys}^a$ and $\phi_{phys}^a$ are {\it real, physical}
fields, with masses $m_a=ms_a(s_a+1)$. In terms of this physical field
decomposition, the cubic term (\ref{cubic}) becomes
\beq
{|v|\over 2\sqrt{2}}\phi_{phys}^a A_{phys}^b A_{phys}^c
\left(\sqrt{s_b(s_b+1)s_c(s_c+1)} \sum_{d=1}^r {\mu^{(a)}_d \mu^{(b)}_d
\mu^{(c)}_d\over \phi_{(0)}^d}\right)
\label{three}
\eeq

In general, these three-point coupling terms are nonzero. Therefore, in the
maximally broken phase the coupled-abelian systems involve interactions between
gauge and scalar fields of different masses. All possible couplings must be
taken into account and the renormalization shifts for given external
$A_{phys}^a$ fields may be computed from (\ref{shift}) using (\ref{exp}) for
the masses and (\ref{three}) for the vertex factors. (Note that the
normalizations are such that the restricted $SU(2)$ case is a single abelian
system.) The shifts $\Delta\kappa$ for $SU(2)$, $SU(3)$, $SU(4)$, and $SU(5)$
are presented in Table \ref{shifts} for all the various inequivalent vacua, and
for the various physical gauge field components $A_{phys}^a$. Certain patterns
are clear from these results. First, in general the renormalization shift is
{\it not} an integer multiple of $\frac{1}{4\pi}$. Second, one can recognize
from the various vacua the contributions corresponding to the maximal embedding
vacuum of the lower subalgebras. The only vacua for which the renormalization
shift {\it is} a multiple of $\frac{1}{4\pi}$ are those which correspond to
disconnected and isolated $SU(2)$ components. In these vacua the model reduces
(at cubic level) to a set of independent abelian models, for each of which the
shift is known to be $\frac{1}{4\pi}$ owing to the degeneracy of the gauge and
Higgs masses, which is in turn due to the self-dual nature of the system.

These results indicate that (unlike in the abelian self-dual theory
\cite{kao2}) the self-duality of the coupled-abelian model is not, in itself,
sufficient to produce integer multiples of $\frac{1}{4\pi}$ for
the one-loop renormalization shift of the Chern-Simons coefficient $\kappa$.
Several aspects deserve to be investigated further. First, it would be
interesting to analyse these models, which in a sense lie between the abelian
and nonabelian cases, using the effective action techniques which have been
applied to abelian Chern-Simons-Higgs theories \cite{khare}. Presumably the
various zero-momentum parity-odd contributions to the gauge self-energy
computed using (\ref{shift}) can be interpreted as coming from a term in the
effective action that mimics a Chern-Simons term at large distance. Second, the
role of fermionic fields in a supersymmetric realization would also be of
interest. Finally, these results should be extended to the fully nonabelian
self-dual system.

\vskip 1cm

\thanks{This work has been supported in part by the D.O.E. through grant number
DE-FG02-92ER40716.00, and by the University of Connecticut Research
Foundation.}

\begin{table}
\center
\begin{tabular}{|c|c|cccc|}\hline
Algebra&Vacuum $\phi_{(0)}$& $A_{phys}^1$ & $A_{phys}^2$ & $A_{phys}^3$ &
$A_{phys}^4$  \\\hline
$SU(2)$&${1 \over \sqrt{2}} E_1$ & 1 &  & &    \\\hline
$SU(3)$&${1 \over \sqrt{2}} E_1$ & 0 & 1 & &    \\
&$E_1+E_2$   & ${1 \over 2}$ & ${13 \over 8}$ & &  \\\hline
$SU(4)$&${1\over \sqrt{2}} E_1$& 0 & 0 & 1 &       \cr
&${1\over \sqrt{2}} E_1+{1\over \sqrt{2}} E_3$& 0 &1 &1 &    \cr
&$E_1 + E_2$ & 0 &${1\over2}$&${13\over8}$&       \cr
&$\sqrt{3\over2}E_1+\sqrt{2}E_2+\sqrt{3\over2}E_3$&${3\over10}$&
${115 \over108}$&${933 \over490}$& \\ \hline
$SU(5)$&${1\over\sqrt{2}} E_1$ & 0 & 0 & 0 & 1    \cr
&${1\over\sqrt{2}}E_1+{1\over\sqrt{2}}E_3$ & 0 & 0 &1 &1   \cr
&${E_1 + E_2}$ & 0 & 0 &${1\over2}$&${13\over8}$ \cr
&${1\over\sqrt{2}}E_1 +E_3+E_4$&0&${1\over2}$& 1 & ${13\over8}$\cr
&$\sqrt{3\over2}E_1+\sqrt{2}E_2+\sqrt{3\over2}E_3$ & 0 & ${3\over10}$ &
${115\over108}$ & ${933\over490}$   \cr
&${\sqrt{2}E_1+\sqrt{3}(E_2+E_3)+\sqrt{2}E_4}$ & ${1\over5}$ &
${4549\over6048}$ & ${118553\over82810}$  &
${1488125\over731808}$ \\\hline
\end{tabular}\\
\caption{The additive renormalization shifts $\Delta\kappa$, in units of
$\frac{1}{4\pi}sign(\kappa)$, for the various inequivalent nontrivial vacua
$\phi_{(0)}$ of $SU(2)$, $SU(3)$, $SU(4)$ and $SU(5)$, for the various physical
components $A_{phys}^a$ of the gauge field.}
\label{shifts}
\end{table}  \b


\clearpage


\begin{thebibliography}{99}

\bibitem{deser}S. Deser, R. Jackiw and S. Templeton, ``Topologically Massive
Gauge Theory'', {\it  Ann. Phys.} {\bf 140} (1982) 372.

\bibitem{pisarski}R.~Pisarski and S.~Rao, ``Topologically Massive
Chromodynamics in the Perturbative Regime'', {\it Phys. Rev.} {\bf D32} (1985)
2081.

\bibitem{hong} J.~Hong, Y.~Kim and P-Y.~Pac, ``Multivortex Solutions of the
Abelian Chern-Simons-Higgs Theory", {\it Phys. Rev. Lett.} {\bf 64} (1990)
2330; R.~Jackiw and E.~Weinberg, ``Self-Dual Chern-Simons Vortices", {\it Phys.
Rev. Lett.} {\bf 64} (1990) 2334; R.~Jackiw, K.~Lee and E.~Weinberg,
``Self-Dual Chern-Simons Solitons", {\it Phys. Rev. D} {\bf 42} (1990) 3488;
G.~Dunne, {\it Self-Dual Chern-Simons Theories} (Springer-Verlag, Berlin,
1995).

\bibitem{klee} K.~Lee, ``Relativistic nonabelian self-dual Chern-Simons
systems '', {\it Phys. Lett. B} {\bf 255} (1991) 381, ``Self-Dual Nonabelian
Chern-Simons Solitons", {\it Phys. Rev. Lett.} {\bf 66} (1991) 553; G.~Dunne,
``Relativistic Self-Dual Chern-Simons Vortices with Adjoint Coupling'', {\it
Phys. Lett. B} {\bf 324} (1994) 359.

\bibitem{kao1} H-C.~Kao and K.~Lee, ``Self-Dual $SU(3)$ Chern-Simons Higgs
Systems'', {\it Phys. Rev. D} {\bf 50} (1994) 6626.

\bibitem{dunne1} G.~Dunne, ``Vacuum Mass Spectra for $SU(N)$ Self-Dual
Chern-Simons-Higgs Systems'', {\it Nucl. Phys. B} {\bf 433} (1995) 333.

\bibitem{khlebnikov}S.~Yu.~Khlebnikov and M.~Shaposhnikov, ``Spontaneous
Symmetry Breaking Versus Spontaneous Parity Violation'', {\it Phys. Lett.} {\bf
 B254} (1991) 148.

\bibitem{khare} A.~Khare, R.~MacKenzie and M.~Paranjape, ``On the Coleman-Hill
Theorem'', {\it Phys. Lett. B} {\bf 343} (1995) 239; H-C.~Kao, ``Generalizing
the Coleman-Hill Theorem'', hep-th/9506093.

\bibitem{lusheng} L.~Chen, G.~Dunne, K.~Haller and E.~Lim-Lombridas, ``Integer
Quantization of the Chern-Simons Coefficient in a Broken Phase'', {\it Phys.
Lett. B} {\bf 348} (1995) 468; A.~Khare, R.~MacKenzie, P.~Panigrahi and
M.~Paranjape, ``Spontaneous Symmetry Breaking and the Renormalization of the
Chern-Simons Term'', {\it Phys. Lett. B} {\bf 355} (1995) 236.

\bibitem{coleman}S.~Coleman and B.~Hill, ``No More Corrections to the
Topological Mass Term in $QED_3$'', {\it Phys. Lett. B} {\bf 159} (1985) 184.

\bibitem{kao2} H-C.~Kao, K.~Lee, C.~Lee and T.~Lee, ``The Chern-Simons
Coefficient in the Higgs Phase'', {\it Phys. Lett. B} {\bf 341} (1994) 181;
H-C.~Kao, K.~Lee and T.~Lee, ``The Chern-Simons Coefficient in Supersymmetric
Yang-Mills-Chern-Simons Theories'', {\it Phys. Lett. B} {\bf 373} (1996) 94.


\bibitem{dunne2} G.~Dunne, ``Mass Degeneracies in Self-Dual Models'', {\it
Phys. Lett. B} {\bf 324} (1994) 359.

\bibitem{dynkin} E.~Dynkin, ``Semisimple Subalgebras of Semisimple Lie
Algebras'', {\it Amer. Math. Soc. Transl.} {\bf 6} (1957) 111;
B.~Kostant, ``The Principal 3-Dimensional Subgroup and the Betti Numbers of a
Complex Simple Lie Group'', {\it Amer. J. Math.} {\bf 81} (1959) 973.

\bibitem{sorba} K.~Bitar and P.~Sorba, ``Classification of Pseudoparticle
Solutions in Gauge Theories'', {\it Phys. Rev. D} {\bf 16} (1977) 431;
A.~Leznov and M.~Saveliev, ``Representation Theory and Integration of Nonlinear
Spherically Symmetric Equations of Gauge Theories'', {\it Commun. Math. Phys.}
{\bf 74} (1980) 111; N.~Ganoulis, P.~Goddard and D.~Olive, ``Self-Dual
Monopoles and Toda Molecules'', {\it Nucl. Phys. B} {\bf 205 [FS]} (1982)
601.

\bibitem{susy} C.~Lee, K.~Lee and E.~Weinberg, ``Supersymmetry and Self-Dual
Chern-Simons Systems'', {\it Phys. Lett. B} {\bf 243} (1990) 105; E.~Ivanov,
``Chern-Simons Matter Systems with Manifest N=2 Supersymmetry'', {\it Phys.
Lett. B} {\bf 268} (1991) 203; S.~J.~Gates and H.~Nishino, ``Remarks on N=2
Supersymmetric Chern-Simons Theories'', {\it Phys. Lett. B} {\bf 281} (1992)
72.

\end{thebibliography}
\end{document}